\documentclass[aps,prl,twocolumn,showpacs,superscriptaddress,groupedaddress]{revtex4}  
\usepackage{graphicx}  
\usepackage{dcolumn}   
\usepackage{bm}        
\usepackage{amssymb}   

\begin{document}

\begin{titlepage}

\title{Quantum Manifestation of Elastic Constants in Nanostructures}

\author{Miao Liu}
\author{Feng Liu\footnote{email: fliu@eng.utah.edu}}

\affiliation{Department of Materials Science and Engineering, University of Utah, Salt Lake City, Utah 84112, USA}

\date{\today}
\begin{abstract}
Generally, there are two distinct effects in modifying the properties of low-dimensional nanostructures: surface effect (SS) due to increased surface-volume ratio and quantum size effect (QSE) due to quantum confinement in reduced dimension. The SS has been widely shown to affect the elastic constants and mechanical properties of nanostructures. Here, using Pb nanofilm and graphene nanoribbon as model systems, we demonstrate the QSE on the elastic constants of nanostructures by first-principles calculations. We show that generally QSE is dominant in affecting the elastic constants of metallic nanostructures while SS is more pronounced in semiconductor and insulator nanostructures. Our findings have broad implications in quantum aspects of nanomechanics.

\end{abstract}

\pacs{62.25.-g, 62.20.-x, 68.60.Bs, 73.22.-f}

\maketitle

\end{titlepage}
Mechanical properties of nanoscale solid structures are different from their bulk counterparts. It has been demonstrated experimentally that elastic moduli change their values as a function of the size of nanostructures, such as the diameter of nanorods or thickness of nanoplates \cite{1,2}. The general understanding is that such size dependence of elasticity has its physical origin in the elasticity variation at material's surface. It is well-known that surface has a different structure from underlying bulk due to bond breaking, surface relaxation and reconstruction \cite{3,4,5}, which gives rise to excess surface energy and non-zero intrinsic surface stress \cite{3,4,5}. Consequently, the elastic constants of surface (which may include several atomic layers \cite{4,5}) are distinctively different from those of bulk. In a nanostructure, the surface-to-volume ratio continues to increase with the decreasing size, so that the overall elastic constants of the nanostructure will exhibit a strong size dependence.

There have been many studies about the elastic constants of nanostructures focusing on the surface effect (SS). For example, experiments showed that Young's modulus of a thin film can either increase or decrease relative to bulk when the film thickness approaches nanoscale \cite{6,7}. Theoretically, it is found that the surface could decrease Young's modulus down to 2/3 of its bulk value from calculations using harmonic or Lennard-Jones potential approximation \cite{8,9}.  Another calculation found that Young's modulus of thin film varies as inverse of its thickness, which could go either larger or smaller than the bulk value, based on embedded atom method (EAM) and Stillinger-Weber potential \cite{10}. EAM simulations of Cu film showed that Cu surface could become either stiffer or softer relative to bulk \cite{11}. In general, elastic constants of nanostructures have been modeled by partitioning the structure into two parts of inner bulk and outer surface with modified surface elastic constants \cite{11,12,13}. This makes the overall mechanical properties of nanostructures distinctively different from those of their bulk counterpart. For example, the mechanical bending of nanofilms follows the modified Stoney \cite{12} and Timoshenko \cite{13} formula rather than the classical formula for macroscopic thick films.

Besides the SS, it is well known that there is another effect becomes increasingly prominent at the nanoscale to affect the properties of low-dimensional nanostructures: quantum size effect (QSE) induced by quantum confinement. When the dimension of a nanostructure is reduced to be comparable to the electron Fermi wavelength, electrons become geometrically confined giving rise to quantized electronic states that change electronic energy, which in turn modify various properties of nanostructures by QSE, such as surface energy \cite{14}, stability \cite{15} and magnetism \cite{16}. Therefore, it is reasonable to expect the QSE to affect the mechanical properties of nanostructures. A few recent theoretical \cite{17,18,19} and experimental \cite{20} studies have indeed shown the QSE causing quantum oscillations of surface (edge) stress in nanostructures. In general, however, despite the extensive study of the SS on the elastic constants of nanostructures \cite{1,2,6,7,8,9,10,11,12,13}, little attention has been paid to the QSE.

In this Letter, we demonstrate quantum manifestations of elastic constants in nanostructures induced by QSE using first-principles calculations. Using Pb nanofilms and graphene nanoribbons (GNRs) as model systems, we show that the Young's modulus and Poisson ratio of nanostructures can display an oscillatory dependence on size, i.e., the thickness of nanofilm and the width of nanoribbon. The main physical origin for such quantum oscillations of elastic constants is the QSE induced oscillation of electron density inside the nanostructure. Because electron Fermi wavelength is much shorter in a metal than in a semiconductor or insulator, generally the QSE dominates over the SS in affecting the elastic constants of metal nanostructures, while the reverse is true for semiconductor and insulator nanostructures. It is also important to point out that previous theoretical studies \cite{8,9,10,11,12,13} used empirical potentials which did not account for electronic effects. Consequently, the QSE on elastic constants will be missed in these earlier studies even if it were present, which calls for the need of first-principles methods.

Our calculations are carried out using the density functional theory method as implemented in the VASP code \cite{21} with the projector augmented wave method \cite{22} and the Perdew-Burke- Ernzerhof exchange-correlation functional \cite{23}.  As shown in Fig. 1, Pb(111) film is modeled by a supercell slab set at the theoretical bulk lattice constant of 5.04\AA\ as the reference of strain-free state \cite{24}. The slabs are separated by a vacuum thickness of 20\AA\ in z-direction, sampled by a 20$\times$20$\times$1 mesh in k-space. GNR is modeled by using similar super cell technique with a vacuum thickness of 20\AA\ in both y and z directions, sampled by a 10$\times$1$\times$1 mesh in k-space. All calculations used a plane-wave cutoff of 1.3 times of default VASP value and the structure is optimized until the atomic forces converged to ~1 meV/\AA. We extracted elastic constants of Yong's modulus ($E$) and Poisson ratio ($\nu$) from calculating stress-strain relations as a function of system size, for which we varied the Pb(111) film thickness from 1 to 30 monolayers (MLs) and the armchair GNR (AGNR) width from 1 to 29 atomic rows.

\begin{figure}
\includegraphics[width=5.0cm]{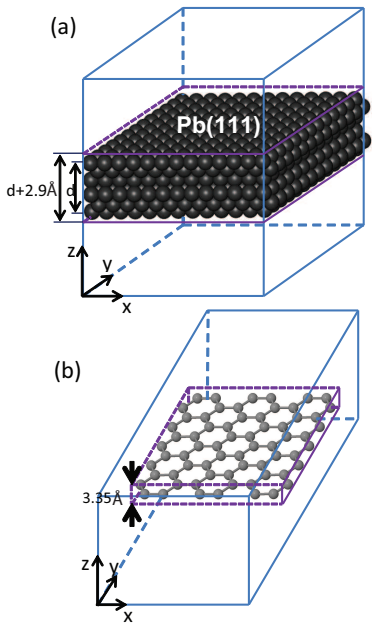}
\caption{\label{fig1} Schematic plot of computational supercell of (a) Pb(111) film and (b) armchair GNR. Vacuum region was shown in outer blue box, and the thickness convention is shown in inner purple box.}
\end{figure}

Young's modulus describes the stiffness of a material, defined as the ratio of tensile stress over tensile strain ($E=\frac{\sigma_\parallel}{\varepsilon_\parallel}$). And Poisson ratio is defined as the ratio between biaxial transverse compressive strain over longitudinal uniaxial tensile strain ($\nu=-\frac{\varepsilon_\perp}{\varepsilon_\parallel}$). Here, for Pb(111) thin film, we apply biaxial compressive strain ($\varepsilon_{x,y}$) in the film surface plane (normal to film surface), and calculate the strain induced film stress in the surface plane ($\sigma_{x,y}$) (note that the intrinsic surface stress in the absence of strain is subtracted) and tensile strain ($\varepsilon_z$) in the surface normal direction. Then we define the film Young's modulus as $E_f=\frac{\Delta\sigma_{x,y}}{\Delta\varepsilon_{x,y}}$ and Poisson ratio as $\nu_f=-\frac{\Delta\varepsilon_{x,y}}{\Delta\varepsilon_z}$, as illustrated in Fig. 1(a).  Similarly, for AGNR, we define $E_r=\frac{\Delta\sigma_x}{\Delta\varepsilon_x}$ and $\nu_r=-\frac{\Delta\varepsilon_y}{\Delta\varepsilon_x}$, as illustrated in Fig. 1(b). Another issue for systems with surface and edge is how to define their thickness and width. Here, we use the convention that the thickness (width) is set equal to the distance between the two outmost atomic planes (rows) plus one interlayer (inter-row) spacing, as show in Fig. 1(a) [Fig. 1(b)]. The interlayer spacing of Pb (111) film is 2.90\AA, and the inter-layer of armchair GNR is 3.35\AA. These same values are used throughout for consistency.

\begin{figure}
\includegraphics[width=6.5cm]{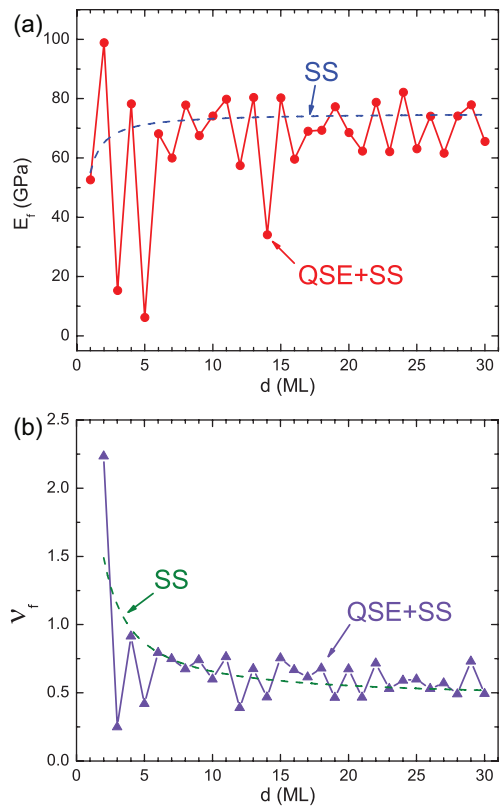}
\caption{\label{fig2} (a) Young's modulus and (b) Poisson ratio of Pb(111) film as a function of film thickness. The dashed lines show the fitted surface effect on Yong's modulus and Poisson ratio.}
\end{figure}

Figure 2(a) and (b) shows the calculated $E_f$ and $\nu_f$ of Pb (111) film as a function film thickness from 1 to 30 MLs, respectively. Clearly, both $E_f$ and $\nu_f$ show a strong odd-even oscillation with a beating pattern period of $\sim$9MLs, manifesting the QSE. Overall, both oscillation patterns are very similar to those of surface energy \cite{25} and surface stress \cite{19} of Pb (111) film. The oscillation amplitude for both $E_f$ and $\nu_f$ decays slowly, remains to be strong even for $\sim$30ML thick film. Apparently, they differ from commonly recognized trend that elastic constants change monotonically as a function of film thickness due to SS \cite{8,9,10,11,12,13}. This clearly demonstrates the importance of QSE on modifying the elastic constants of nanofilms.

To better understand the physical origin of quantum oscillations of elastic constants, we may consider a simple free electron gas model. The bulk modulus of a uniform electron gas of density $n$ is \cite{26}
\begin{equation}
\label{Eq_1}
 B(n)=\frac{\hbar^2(3\pi^2)^{2/3}}{3m}n^{5/3}\propto n^{5/3}
\end{equation}

\begin{figure}
\includegraphics[width=8.0cm]{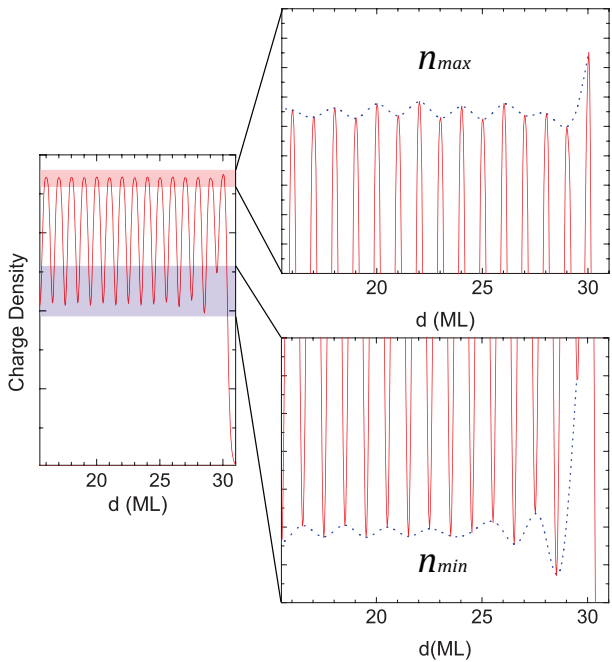}
\caption{\label{fig3} Charge density distribution along z direction of 30ML Pb(111) film. (a) $n_max$ show the maximum charge density within each atomic planes; (b) $n_min$ show the minimum charge density in between atomic planes.}
\end{figure}

In a nanofilm, QSE modulates the electron density along the surface normal z-direction. Figure 3 shows the charge density distribution $n(z)$ along the z-direction in the 30ML film. Clearly, both the maximum charge density within an atomic plane ($n_max$) and the minimum density in between two atomic planes ($n_min$) exhibit an odd-even periodic oscillation originated from the QSE along the z-direction of the film. Thus, approximately, the elastic modulus of the film can be calculated as
\begin{equation}
\label{Eq_2}
 B_f=\frac{1}{d}\int\limits_0^d B(n,z)dz\propto\frac{1}{d}\int\limits_0^d n(z)^{5/3}dz
\end{equation}
And the QSE modulated charge density distribution leads to the QSE modulated film modulus and similarly other elastic constants.

Besides the QSE, the SS should be present also. If one considers the film has an elastic constants ($C_b$), thickness ($d$) and a surface layer of thickness ($\delta$) and surface elastic constants ($C_s$) \cite{12,13}, the overall film elastic constant can be easily calculated as
\begin{equation}
\label{Eq_3}
C_f=C_b+\frac{2(C_s-C_b)\delta}{d}
\end{equation}

which shows an inverse linear dependence on the film thickness ($\sim1/d$) cite{11,12,13}. Whether the surface becomes harder or softer depends on $C_s$. If $C_s>C_b$, $C_f$ increases with decreasing $d$; if $C_s<C_b$, $C_f$ decreases with decreasing $d$. Now, if we pretend to ignore the QSE and use Eq. (3) to forcefully fit the calculated the $E_f$ and $\nu_f$, we got the black dashed lines shown in Fig. 2, which in fact reflect the SS. From the fitting, we obtained $E_b\sim$75 GPa and $\nu_b\sim$0.5 at the bulk limit, which are in good agreement with experimental values of 80 GPa and $\nu$=0.4 \cite{27}.

\begin{figure}
\includegraphics[width=6.5cm]{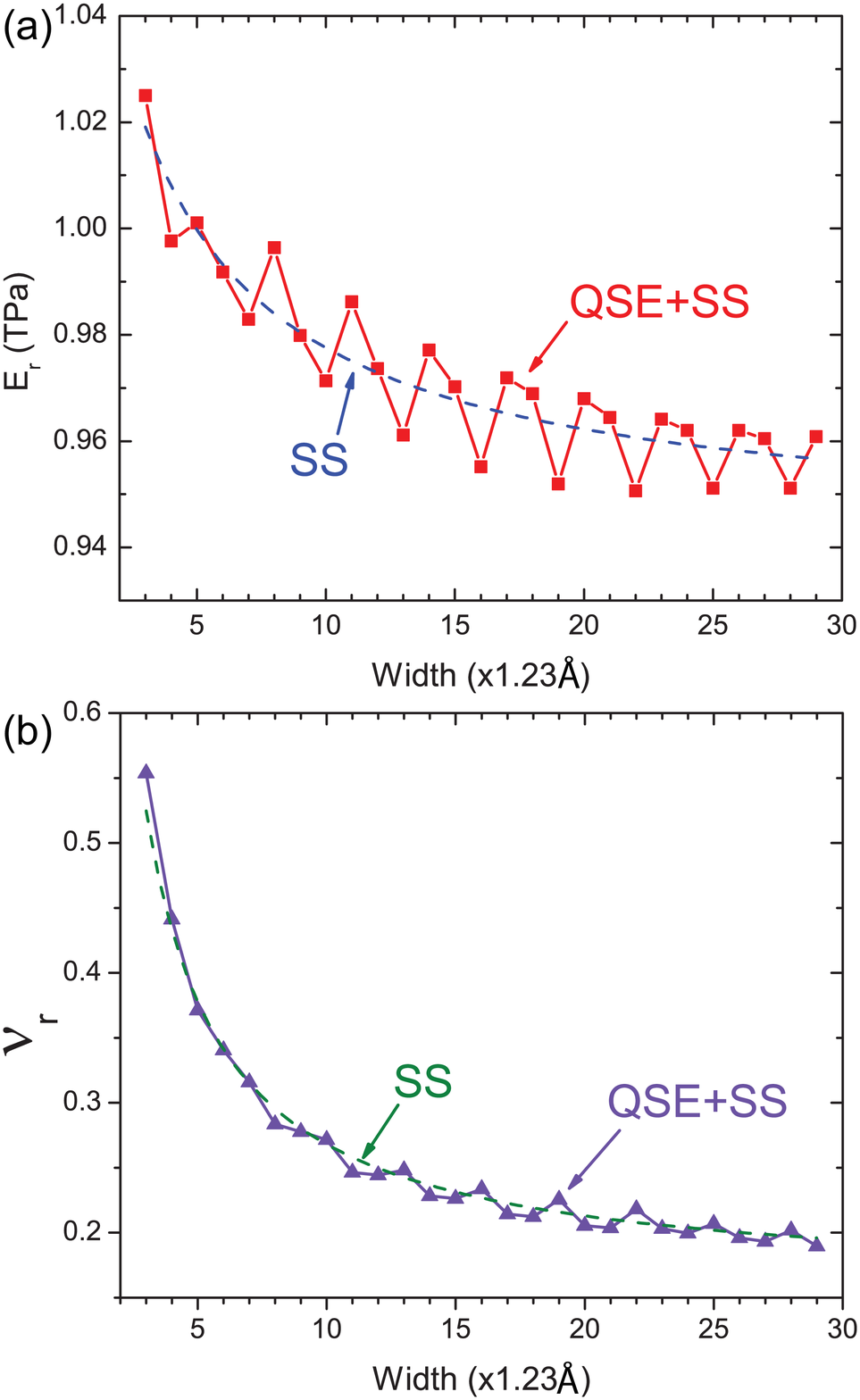}
\caption{\label{fig4} (a) Young's modulus and (b) Poisson ratio of AGNR as a function of ribbon width.  The dashed lines show the fitted surface effect on Young's modulus and Poisson ratio.}
\end{figure}

For comparison, we also preformed similar calculations for AGNRs. The reason for choosing AGNR is because it has been known that the AGNR exhibits interesting QSE effect in electron band structure \cite{28}, with 1/3 being metallic and 2/3 being semiconducting in a three-atomic-raw oscillation as a function of width, as well as similar oscillations in edge energy and edge stress \cite{17} (i.e., equivalent surface energy and stress in 2D). Figure 4 shows the calculated Young's modulus ($E_r$) and Poisson ratio ($\nu_r$) of AGNR as a function of width from 1 to 29 atomic rows. Not too surprising, we see the similar quantum oscillations in both $E_r$ and $\nu_r$ induced by QSE, with a three-atomic-raw period as in edge energy and edge stress \cite{17}. In addition, we also fit the data using Eq. (3) to reveal the edge effect (i.e., the equivalent SS in 2D), shown as the dashed line in Fig. 4. From the fitting, we obtained the graphene Young's modulus and Poisson ratio to be 0.95 TPa and 0.16, in good agreement with previous works \cite{29,30}.

It is interesting to compare the results of Pb(111) film in Fig. 2 with those of AGNR in Fig. 4, to reveal the relative importance of QSE versus SS. For Pb (111) film, the QSE modulation of elastic constant is so strong with an oscillation magnitude changing the elastic constants by $\sim$100\%, while the SS is less important changing the elastic constant by at most $\sim$26\%. In contract, for AGNR, the QSE induced oscillation magnitude is very small changing the Young's modulus by a maximum of 2\% and Poisson ratio by 11\%, respectively, while the edge effect (the equivalent SS in 2D) changes them by as much as 7\% and 340\%, respectively. Therefore, we conclude that the QSE dominates over the SS in affecting the elastic constants of Pb nanofilms, while the SS dominates over the QSE in affecting the elastic constants of AGNRs. This can be generally understood as follows. First, consider a free electron gas model, the electron Fermi wavelength ($\lambda_F$) scales inversely with electron density ($n$) in a power-law ($\lambda_F\propto n^{-1/3}$ for 3D electron gas and $\lambda_F\propto n^{-1/2}$ for 2D electron gas)\cite{26}, so that the Fermi wavelength is usually shorter in metals with a high electron density than in semiconductors and insulators with a low carrier density. Consequently, quantum confinement of electrons or carries, and hence the QSE is stronger in metal nanostructures than in semiconductor and insulator nanostructures. Second, the metal surfaces usually relax or reconstruct less than the semiconductor and insulator surfaces \cite{3}, so that the SS is expected to be weaker in metal nanostructures than in semiconductor and insulator nanostructures.

In summary, we have demonstrated that the QSE can have a profound effect in affecting the elastic constants of nanostructures, with interesting manifestation of size-dependent quantum oscillations, in addition to the monotonic size-dependence induced by SS that has been widely recognized before. Most importantly, we show that for metal nanostructures the QSE induced oscillations can be the most dominant effect to completely overwhelm the SS. Our findings shed important new light to our understanding of the mechanical properties of nanostructures by adding interesting quantum aspects to nanomechanics with broad implications.

This work was support by NSF-MWN and Materials Theory program (Grant No. DMR-0909212). We also thank DOE-NERSC and the CHPC at University of Utah for providing the computing resources.

\newpage


\begin{references}
\bibitem{24} Following the original Gibbs' definition of \textquotedblleft excess\textquotedblright\ surface quantity of the dividing surface of a bulk [1], surface quantities, including those of nanostructures, should always be defined relative to bulk. For a nanofilm, one may image the film sitting on a substrate. Note that if one fully relaxes the film structure, surface quantities would become ill-defined.
\end{references}
\end{document}